%% file: templateArxiv.tex
\documentclass{article}
\usepackage{PRIMEarxiv}
\usepackage[utf8]{inputenc} 
\usepackage[T1]{fontenc}    
\usepackage{url}            
\usepackage{booktabs}       
\usepackage{amsfonts}       
\usepackage{nicefrac}       
\usepackage{microtype}      
\usepackage{lipsum}
\usepackage{fancyhdr}       
\usepackage{graphicx}       
\graphicspath{{media/}}     
\usepackage{amsmath}
\usepackage{listingsutf8}
\usepackage{pmboxdraw}
\usepackage{float}
\usepackage{placeins}
\usepackage{subcaption}
\usepackage{hyperref}       

\title{A Review of Several Keystroke Dynamics Methods}

\author{
  Soykat Amin \\
  Sapienza Università di Roma \\
  \texttt{amin.1985500@studenti.uniroma1.it} \\
   \And
  Cristian Di Iorio \\
  Sapienza Università di Roma \\
  \texttt{diiorio.1983177@studenti.uniroma1.it} \\
}

\begin{document}
\maketitle

\begin{abstract}
Keystroke dynamics is a behavioral biometric that captures an individual’s typing patterns for authentication and security applications. This paper presents a comparative analysis of keystroke authentication models using Gaussian Mixture Models (GMM), Mahalanobis Distance-based Classification, and Gunetti Picardi’s Distance Metrics. These models leverage keystroke timing features such as hold time (H), up-down time (UD), and down-down time (DD) extracted from datasets including Aalto, Buffalo and Nanglae-Bhattarakosol. Each model is trained and validated using structured methodologies, with performance evaluated through False Acceptance Rate (FAR), False Rejection Rate (FRR), and Equal Error Rate (EER). The results, visualized through Receiver Operating Characteristic (ROC) curves, highlight the relative strengths and weaknesses of each approach in distinguishing genuine users from impostors.
\end{abstract}


\include{chapters/ch1}
\include{chapters/ch2}
\include{chapters/ch3}
\include{chapters/ch4}
\include{chapters/ch5}

\include{bibliography}
\end{document}

%% file: chapters/ch1.tex
\section{Introduction to Keystroke Recognition}
Authentication methods follow three categories: something you \textit{know}, like passwords, something you \textit{have}, like tokens and something you \textit{are}, like biometrics \cite{ref:glasgow intro}. Biometrics can be divided into two categories, physiological biometrics and behavioral biometrics. The former refers to characteristics of the human body that do not change (face, iris, fingerprint) while the latter involves using a person's action to distinguish them.\\

Keystroke dynamics falls into the behavioral biometrics category. It works by measuring and assessing people’s typing rhythm on digital devices like computer keyboards or mobile phones\cite{ref:survey intro}.\\

\subsection{Advantages}
Keystroke Recognition has numerous advantages, like:
\begin{itemize}
    \item High acceptability and collectability. It does not require specialized hardware unlike many other biometric systems such as fingerprint recognition and iris recognition. The only physical component needed is a standard computer keyboard. This means that they can even be deployed remotely.
    \item Uniqueness. Modern computer keyboard software can measure events with high precision, meaning that it is nearly impossible to replicate another user's typing rythm.
    \item Universality. Almost everyone uses a keyboard for various activities.
    \item Low cost. Because there is no need for specialized hardware, the costs are low.
\end{itemize}

\subsection{Disadvantages}
However there are some disadvantages:
\begin{itemize}
    \item Low accuracy. This kind of biometric system is generally worse than other systems like iris recognition or face recognition.
    \item Low permanence. Since this is a behavioral biometric, it has lower permanency than physiological biometrics since typing patterns can change with time.
\end{itemize}

\subsection{Timing Features and Data Types}
The timing features of keystrokes are latency and hold time \cite{ref:air force intro}. Latency is the time between consecutive key presses/releases while hold time is the time between key press and key release. This means that between two consecutive keystrokes there are 6 timing features that can be analyzed: hold time of two keys, time between key presses, time between key releases, time between first key press and second key release, time between first key release and second key press.\\

\begin{figure}[h]
    \centering
    \includegraphics[width=0.75\linewidth]{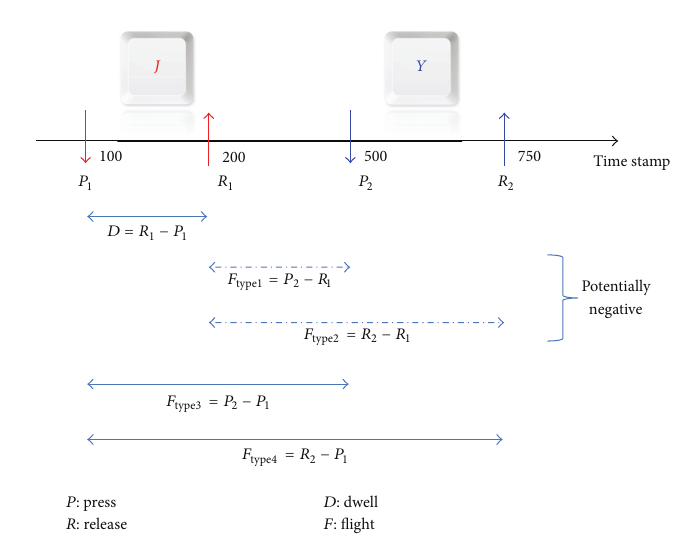}
    \caption{Relevant keystroke events}
\end{figure}

These typing features are extracted from keystroke data, which can be divided into two types:
\begin{itemize}
    \item free-text, when the user can write freely. It is more complex to implement and it suffers from inconsistencies in user input. Those characteristics make it more secure and suitable for continuous authentication.
    \item fixed-text, when the user must type a predefined string. This means that comparison is easier to implement and more precise. However it has limited usability.
\end{itemize}

%% file: chapters/ch2.tex
\section{Datasets}
A keystroke dataset is essential for effectively evaluating keystroke dynamics-based authentication algorithms. The dataset's quality, size and type (fixed or free-text) play a critical role in assessing algorithm performance. After some consideration we chose three datasets \textbf{Aalto 136M}\cite{ref:aalto-136m}, \textbf{Buffalo} \cite{ref:buffalo}, and \textbf{Nanglae-Bhattarakosol} \cite{ref:2dataset}.\\

\subsection{Aalto}

Aaalto 136M dataset includes data from 168.960 participants. It is fixed-text since each participant completed one session which consisted of typing 15 English sentences drawn from a corpus. The 15 sentences were randomly taken from a set of 1.525 sentences consisting of at least 3 words and a maximum of 70 characters. They were asked to type as fast and accurate as possible. Subjects were allowed to make typing errors, correct them or add new characters when typing, and as a result, they could type more than 70 characters. The dataset is categorized as a controlled fixed-text dataset because it involves transcribing - subjects did not type contents of their own but were shown what to type. There are around 810 keystrokes per user which translates to a total of 136.857.600 keystrokes in total.

\subsection{Buffalo}

Buffalo contains both free-text and fixed-text keystrokes from 157 participants. The data was collected in three sessions spread out over 4 months. On average there was a 28 days gap between 2 sessions to consider the temporal variations. The users had two tasks:

\begin{itemize}
    \item  The first task was copying the 2005 commencement speech of Steve Jobs at Stanford University. The speech was divided into 3 parts with each part of equal length, the subjects had to type one part in each session (fixed text);
    \item The second task consisted of multiple subtasks including answering survey questions and describing a picture, sending an email with attachment and free internet browsing (free text).
\end{itemize}

Each session lasted for about 50 minutes with 30 minutes for the first task and 20 minutes for the second task. In one section of the dataset the same keyboard was used across sessions, while in the other section subjects used different keyboards. On average there are around 17000 keystrokes for each subject. Some participant's data had to be discarded due to mistakes, so the dataset contains 148 users instead of 157.\\

\subsection{Nanglae-Bhattarakosol}

Nanglae-Bhattarakosol \cite{ref:nanglae} contains fixed-text data coming from 108 users. They entered static information (name, last name, email address, phone number) 10 times per entry using an iPhone's touchscreen.\\

\subsection{Considerations on datasets}
We chose these three different datasets because they have different characteristics: Aalto has by far the most users and the most data, however it was collected in a single session; meanwhile Buffalo has less data but it's more interesting since it was collected over three sessions. Nanglae-Bhattarakosol was chosen because of its small size.

%% file: chapters/ch3.tex
\section{Authentication Algorithms}
There are several algorithms for keystroke recognition but they can be divided into broad categories: statistical, machine learning based and deep learning based.\\

Statistical algorithms are the easiest to implement as they are deterministic and require low computing power. However, they necessitate manual feature extraction. We chose this type of algorithm for our project for these reasons, as well as due to our limited experience with machine learning and deep learning. The following section will provide a detailed description of the algorithms we used.

\subsection{Mahalanobis Distance}
Mahalanobis distance takes into consideration the covariance of data variables, this makes it better suited for real-world data \cite{ref:mahalanobis}.\\

Euclidean and Manhattan distances have been used for their simplicity. However they are very sensitive to scale variations in the variables and they have no way to deal with correlation between feature variables like Mahalanobis can. The squared Mahalanobis distance between two vectors x and y is defined as:
\begin{equation}
    ||x-y||^2=(x-y)^TS^{-1}(x-y)
\end{equation}
where S is the data's covariance matrix.\\

As you can see, it weights the distance calculation according to the statistical variation of each component and it also decouples interactions between features based on their covariance.

\subsection{Gunetti Picardi}
\label{sec:gp}
This algorithm \cite{ref:gunetti picardi} works best with free text. It evaluates the timing characteristics between kinds of keystrokes:
\begin{itemize}
    \item digraphs, two consecutive keystrokes
    \item trigraphs, three consecutive keystrokes
    \item n-graphs, n consecutive keystrokes
\end{itemize}
Two primary measures are used, which return a value between 0 and 1; they are called R and A.\\

The R (relative) measure, measures the degree of disorder between two samples. To compute the R measure of an array S, you calculate the sum of distances between the position of each element in S and the position of the same element in array S' (its ordered equivalent). The result is a value that falls between 0 and 1 where 0 means no disorder while 1 means maximum disorder. $R_n$ measure the degree of disorder of two samples that shares some n-graphs. $R_{n,m}$ measure the degree of disorder of two samples that shares some n-graphs and some m-graphs. This can be extended to a larger number of different graphs.\\

However the R measure does not consider absolute typing speeds at all, meaning that we need another measure, A.\\

The A (absolute) measure, considers the absolute value of the typing speed of each pair of identical n-graphs in the two samples involved. The A distance between two samples S1 and S2 is:
\begin{equation}
    A_{t,n} = 1- \frac{\text{number of similar n-graphs between S1 and S2}}{\text{total number of n-graphs shared by S1 and S2}}
\end{equation}

Just like R, A can be evaluated through the use of multiple graphical representations, allowing for a more comprehensive analysis. By leveraging two or more distinct graphs, it becomes possible to capture different perspectives on the data, each providing unique insights into the nature of A. For example, the measure $R_{23}A_{23}$ represents a combined approach where both $R_{23}$ and $RA_{23}$ contribute to assessing the similarity or difference between two typing samples. Here, $R_{23}$ is a relative measure that ranks n-graphs (such as digraphs and trigraphs) based on their typing speed, while $A_{23}$ is an absolute measure that evaluates whether the typing durations of these n-graphs fall within an acceptable threshold of similarity. The combination of these measurements enhances accuracy, it also allows for a more robust performance evaluation. When used together, the complementary aspects of these graphs make performance better.

\subsection{Gaussian Mixture Model}
Gaussian Mixture Model (GMM), a linear non-Gaussian multivariate statistical method, is a popular algorithm used for handling non-Gaussian data. It is a statistical method based on the weighted sum of probability density functions of multiple Gaussian distributions \cite{ref:gmm}. GMM generates a vector of mean values corresponding to each component and a matrix of covariance which includes components’ variances and the co-variances between each other. GMM can represent data in higher dimensions than Pure Gaussian by using a discrete set of Gaussian functions, each with its own mean and covariance matrix.\\

GMM is expressed by the parameter set $\lambda$: 
\begin{equation}
    \lambda = \{w_i,\vec{u}_i,\Sigma_i\}, \text{ with } i =1,...,M
\end{equation}
In that equation, $w_i$ are the component weights, $\vec{u}_i$ is the mean vector and $\Sigma_i$ is the covariance matrix.

%% file: chapters/ch4.tex
\section{Feature extraction}

First of all the raw files contained in the datasets need to be processed. This step was entirely developed by us.\\

For Aalto 136M, a raw file in the Aalto dataset contains:
\begin{itemize}
    \item PARTICIPANT\textunderscore ID
    \item TEST\textunderscore SECTION\textunderscore ID
    \item SENTENCE
    \item USER\textunderscore INPUT
    \item KEYSTROKE\textunderscore ID
    \item PRESS\textunderscore TIME
    \item RELEASE\textunderscore TIME
    \item LETTER
    \item KEYCODE
\end{itemize}
We compute Hold Time (H), which measures how long a key is pressed, Up-Down Time (UD), which represents the time between releasing one key and pressing the next, and Down-Down Time (DD), which measures the time between pressing two consecutive keys. Those values are then filtered to remove negative or invalid entries. The function then assigns a subject column based on the user ID and determines the key column using either the LETTER or KEYCODE values. The final dataset consists of selected columns \lstinline|subject, key, H, UD, DD| with missing values removed. The data is merged and exported as a \textit{.csv} file.\\

In the case of the Buffalo dataset, in the raw files each line corresponds to an event, for example:
\begin{verbatim}
    A KeyDown 63578429792961
    A KeyUp 63578429793054
    M KeyDown 63578429793257
    M KeyUp 63578429793382
\end{verbatim}
The first element is the name of the key. The second element is the key event (key down or key up). The third element is the time stamp in milliseconds. From the raw files we calculate Hold time(H), Down-Down time(DD) and Up-Down time(UD) and we save it to a \textit{.csv} file. To do this, we use helper functions stored in \lstinline|utils.py| to traverse all dataset files, save the metadata (like user ID, session, task, ...) and then we read the raw keystroke logs to extract the aforementioned keystroke features while also tracking the number of keystroke repetitions per user. The dataset is divided in two parts, free-text (\lstinline|task==1|) and fixed-text (\lstinline|task==0|). The end result is a \textit{.csv} file containing \lstinline|subject,key,H,UD,DD|.\\

In the case of the Nanglae-Bhattarakosol dataset, the features are already extracted. We made some changes, mainly concerning timings and file structure. We converted the timings from milliseconds to seconds and we merged the three separate \textit{.xlsx} files into a single \textit{.xlsx} file and then we converted it to a \textit{.csv} file.\\

After processing the data from one of the three datasets, we use either: Mahalanobis, Gaussian Mixture Model or Gunetti-Picardi.\\

\lstinline|MahalanobisDetector|. It takes a list of user IDs and a pandas dataframe that contains keystroke timing data. Then mean feature vector for a user's training data is computed; this represents a user's typical behavior. Afterwards, the performance is evaluated using Hold (H), Up-Down (UD) and Down-Down (DD) times. This is done by extracting genuine user keystroke data from the subject and impostor data from other users. Then the genuine data is split 70/30, where the first 70\% is used for training and the remaining 30\% for testing. The first 5 samples from other users are used for impostor testing. To do complete testing, we compute the pseudo-inverse of the covariance matrix between keystroke features. Then we use it to compute the Mahalanobis distance between genuine values and the mean vector. We repeat the same process for impostor values. Finally, the results are used to plot the ROC curve.\\

\lstinline|GMMKeystrokeModel|. The extracted features are then split in: training set (70\%), validation set (30\%). Then for each user and each digraph a Gaussian Mixture Model is fit on the hold time; the means, covariances and weights of the Gaussian distributions are stored. Then the scores are calculated for genuine and impostor values. Performance metrics, like FAR and FRR are calculated while sweeping through each similarity threshold (from 0.0 to 1.0).\\

\lstinline|GunettiPicardi|. After extracting the features, we create a user profile using the extracted features like digraphs, trigraphs, fourgraphs, average keystroke durations and relative timing relationships. This profile is stored in a pickle (\textit{.pkl}) file. Afterwards the system compares an input keystroke sequence against those stored user profiles using Gunetti Picardi distance metrics \ref{sec:gp}: absolute distance (A), which keeps track of timing; Relative distance (R) which measures the difference between samples; a combination of A and R. The closest matching user profile is found and users are classified as either genuine or impostors. Authentication is performed only whether the classified user passes a threshold comparison. To minimize false rejections a secondary check is performed.

%% file: chapters/ch5.tex
\section{Evaluation}
\subsection{Metrics}
The metrics we used are:
\begin{itemize}
    \item FAR, False Acceptance Rate, is the percentage of operations where an impostor claims an identity and a false acceptance occurs.
    \item FRR, False Rejection Rate, is the percentage of operations where a genuine claim is falsely rejected.
    \item EER, Equal Error Rate, which is the error rate where FAR = FRR.
    \item ROC, Receiver Operating Curve, which shows the probability of Genuine Accept (1 - FRR) against the False Accept Rate variation.
\end{itemize}

Due to the fact that keystroke dynamics is behavioral, it is less robust. But it can be useful in an intrusion detection system. That is why we care more about FAR than FRR. The evaluation for each method is done for:

\begin{itemize}
    \item Aalto dataset;
    \item Buffalo free-text dataset;
    \item Buffalo fixed-text dataset;
    \item Nanglae-Bhattarakosol dataset.
\end{itemize}

\subsection{Mahalanobis}
\label{sec:mahalanobis}
To implement this method we used this pre-existing implementation \cite{ref:maha-code} as a template.\\

As you can see from the Figure 2, 3, 4 and 5, Mahalanobis generally has modest authentication performance. On all datasets it has an EER of around 0.35 and the FAR and FRR plots show that it is very sensitive to changes in the threshold, with a steep drop-off as the threshold increases.

\newpage
\begin{figure}[H]
    \centering
    \includegraphics[width=0.7\linewidth]{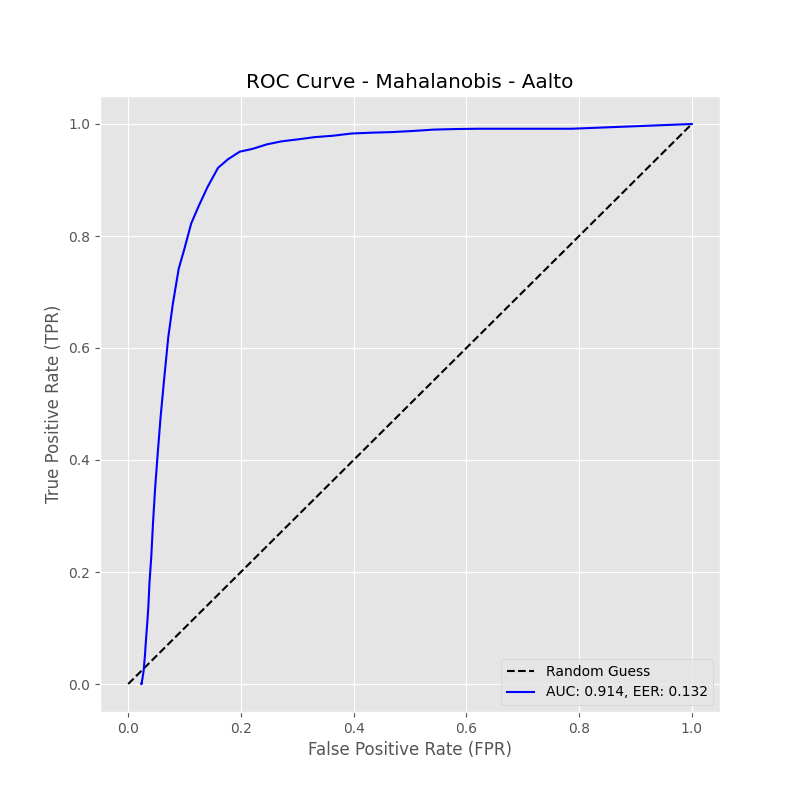}
\end{figure}
\begin{figure}[H]
    \centering
    \begin{minipage}{0.49\textwidth} 
        \centering
        \includegraphics[width=1.2\linewidth]{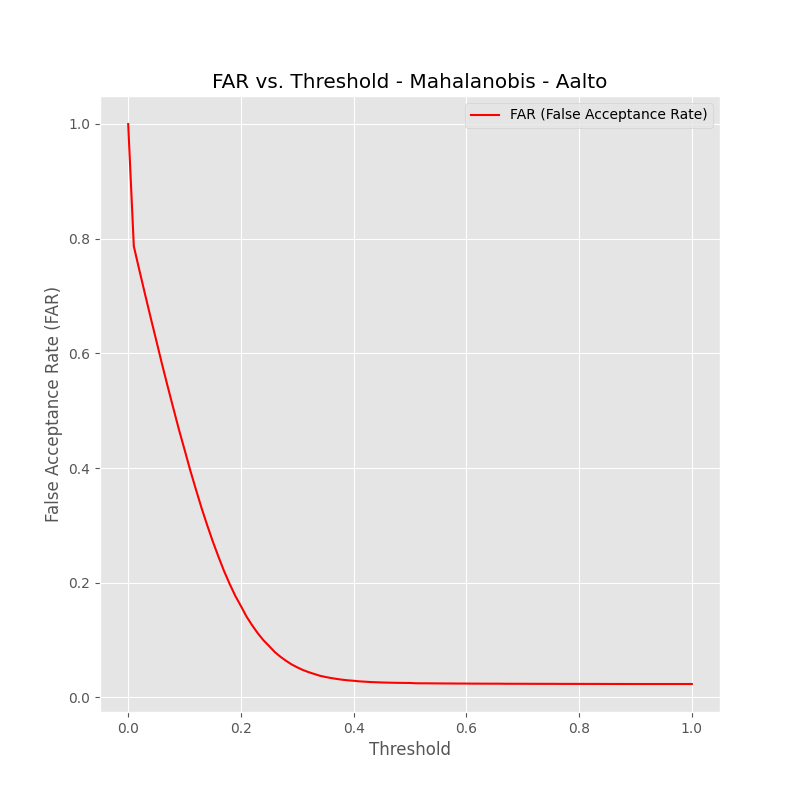}
    \end{minipage}
    \hfill
    \begin{minipage}{0.49\textwidth} 
        \centering
        \includegraphics[width=1.2\linewidth]{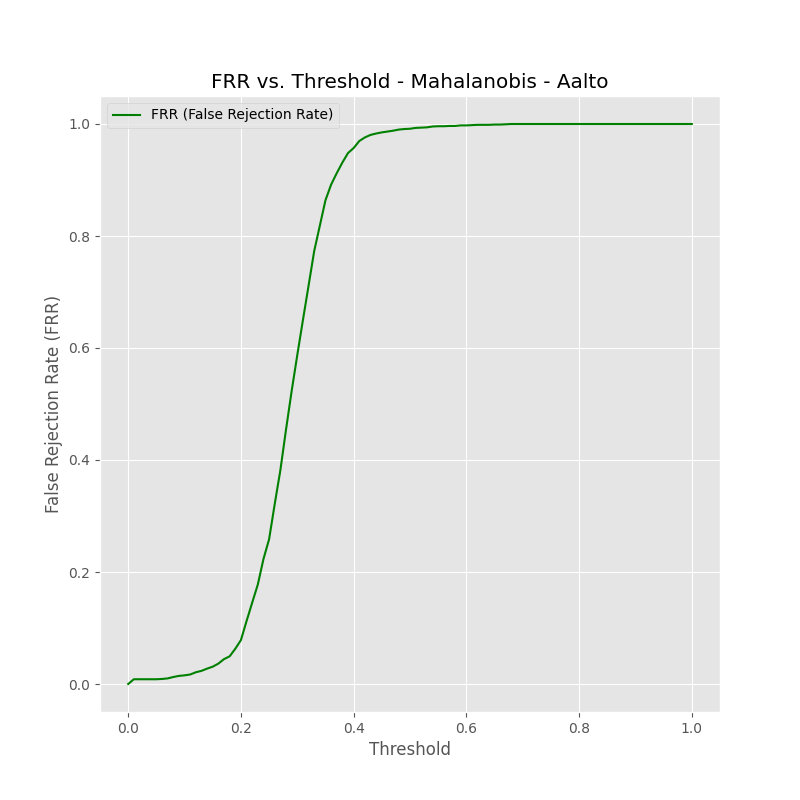}
    \end{minipage}
    \caption{\textbf{Aalto } AUC = 0.914, EER = 0.132}
\end{figure}

\begin{figure}[H]
    \centering
    \includegraphics[width=0.7\linewidth]{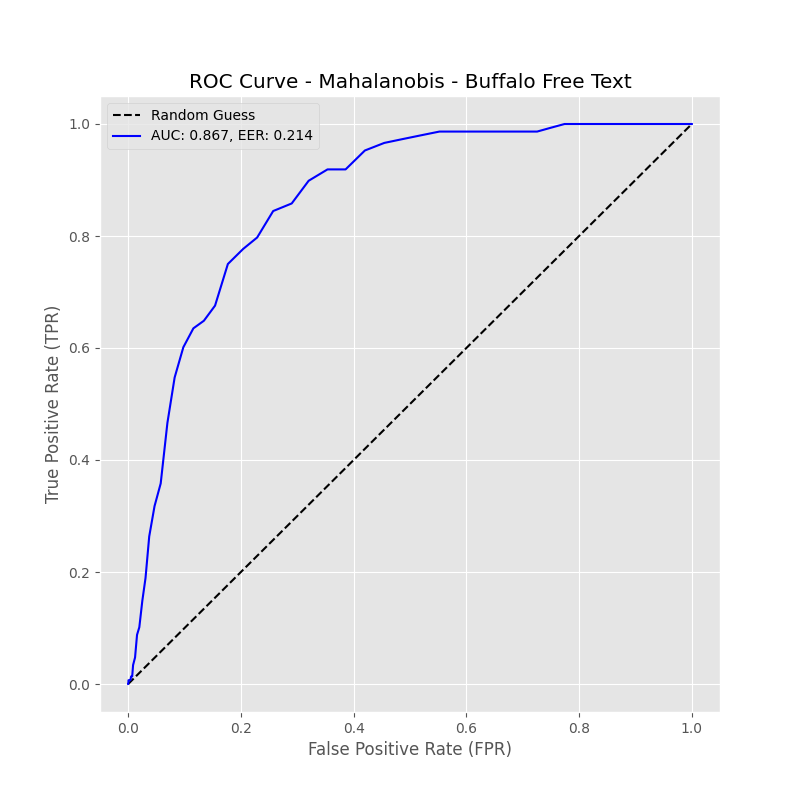}
\end{figure}
\begin{figure}[H]
    \centering
    \begin{minipage}{0.49\textwidth} 
        \centering
        \includegraphics[width=1.2\linewidth]{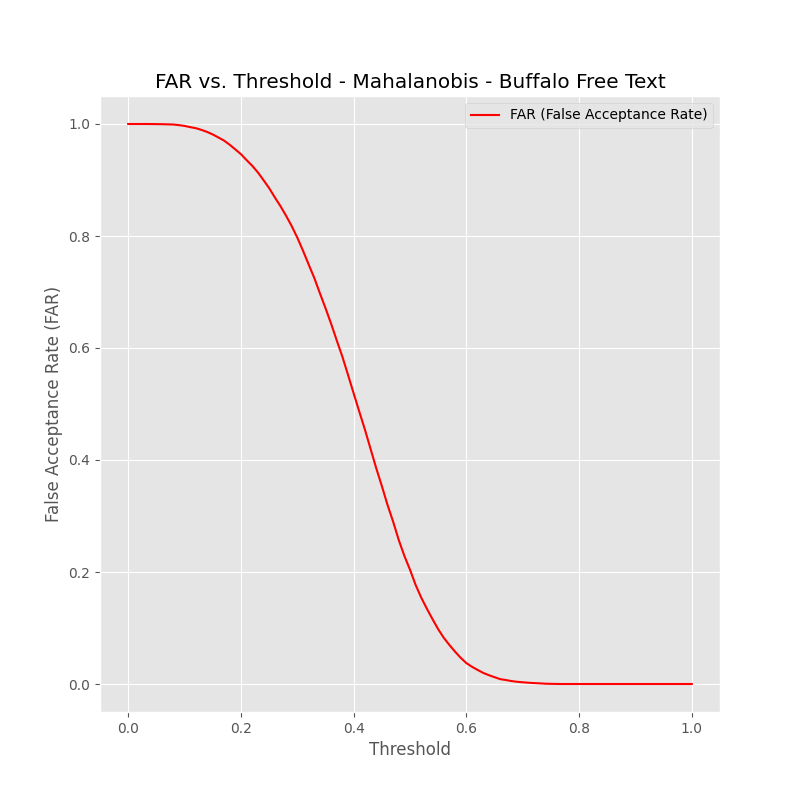}
    \end{minipage}
    \hfill
    \begin{minipage}{0.49\textwidth} 
        \centering
        \includegraphics[width=1.2\linewidth]{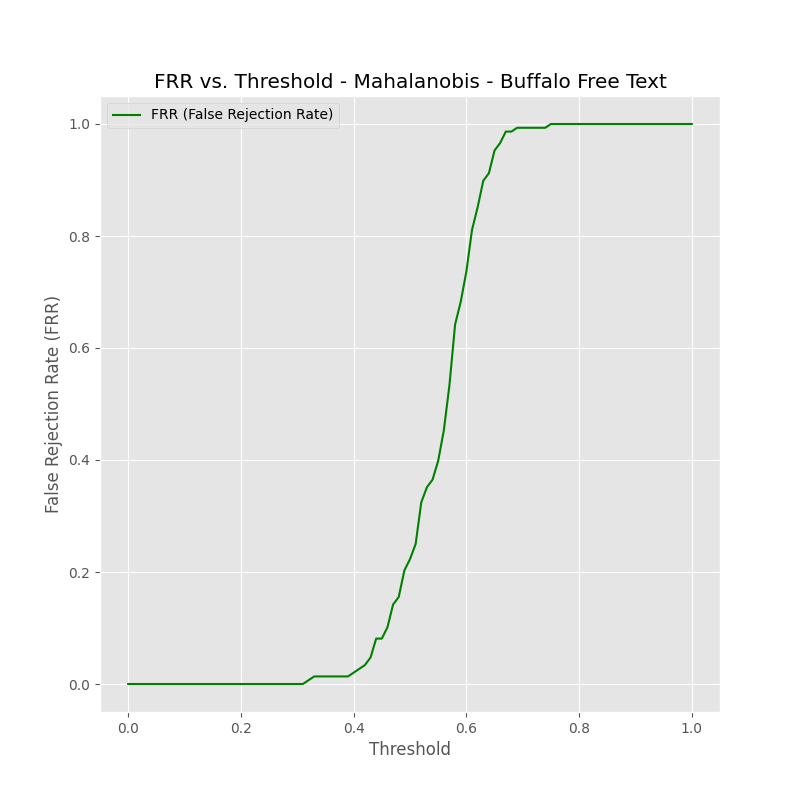}
    \end{minipage}
    \caption{\textbf{Buffalo Free-Text } AUC = 0.867, EER = 0.214}
\end{figure}

\begin{figure}[H]
    \centering
    \includegraphics[width=0.7\linewidth]{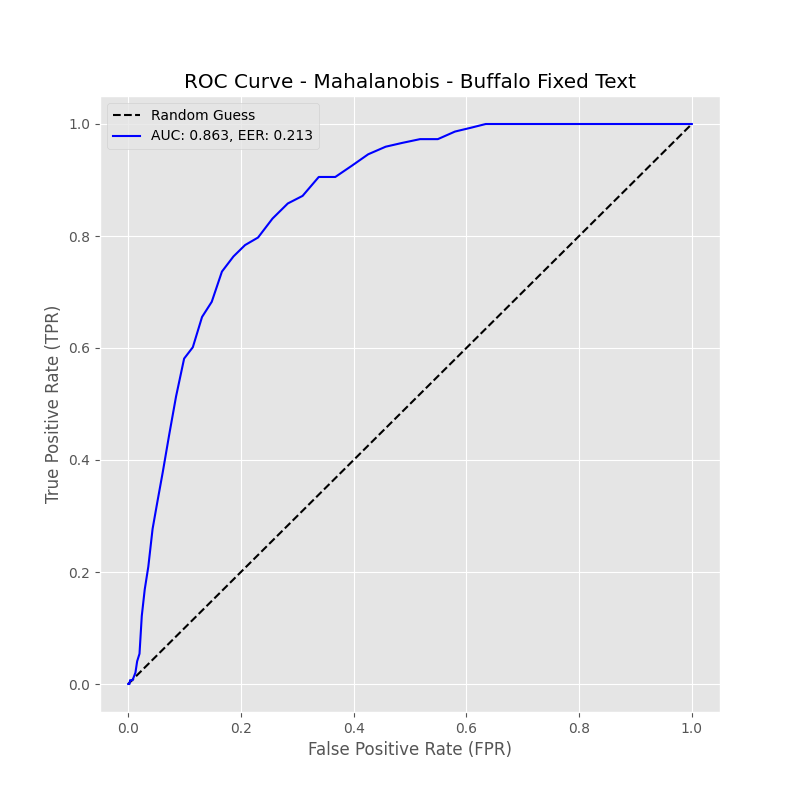}
\end{figure}
\begin{figure}[H]
    \centering
    \begin{minipage}{0.49\textwidth} 
        \centering
        \includegraphics[width=1.2\linewidth]{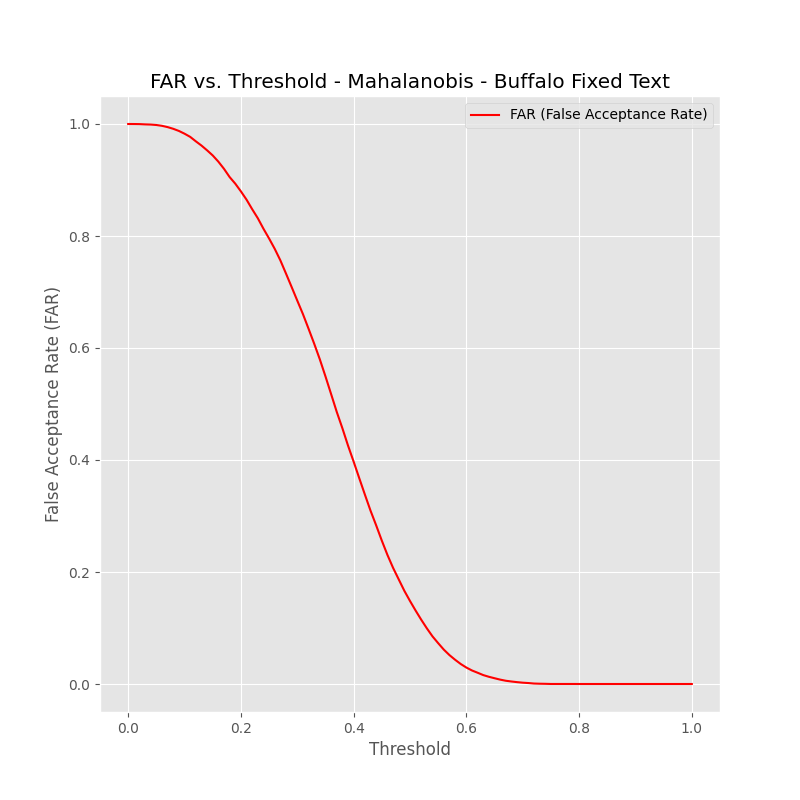}
    \end{minipage}
    \hfill
    \begin{minipage}{0.49\textwidth} 
        \centering
        \includegraphics[width=1.2\linewidth]{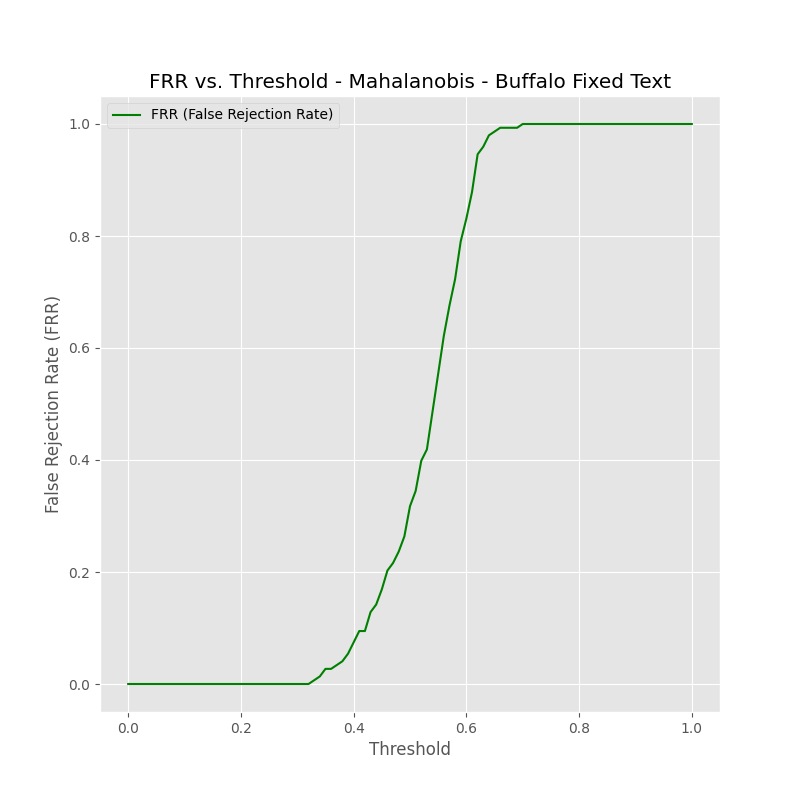}
    \end{minipage}
    \caption{\textbf{Buffalo Fixed-Text } AUC = 0.863, EER = 0.213}
\end{figure}

\begin{figure}[H]
    \centering
    \includegraphics[width=0.7\linewidth]{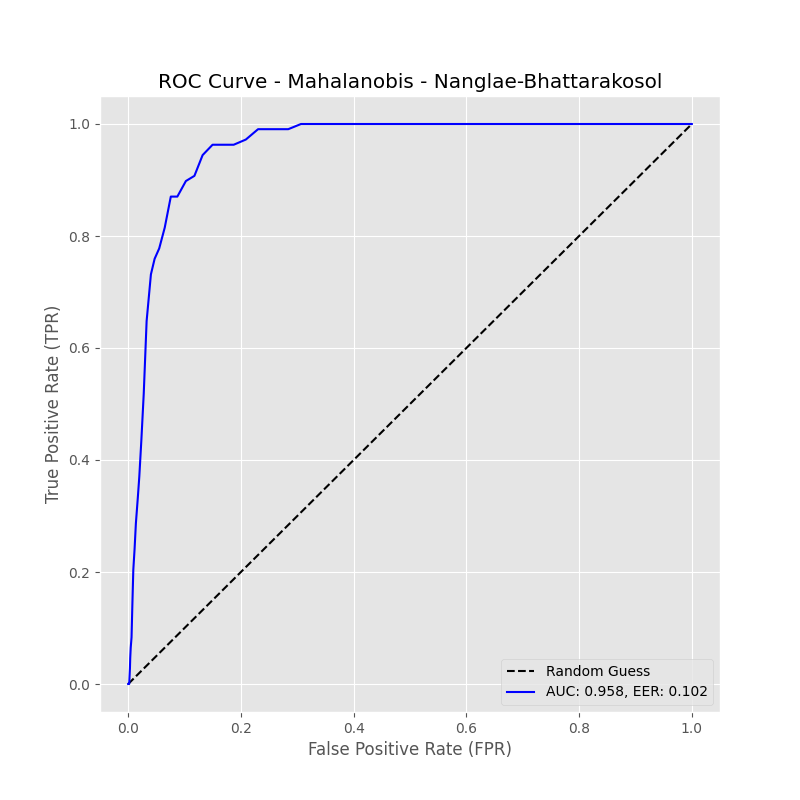}
\end{figure}
\begin{figure}[H]
    \centering
    \begin{minipage}{0.49\textwidth} 
        \centering
        \includegraphics[width=1.2\linewidth]{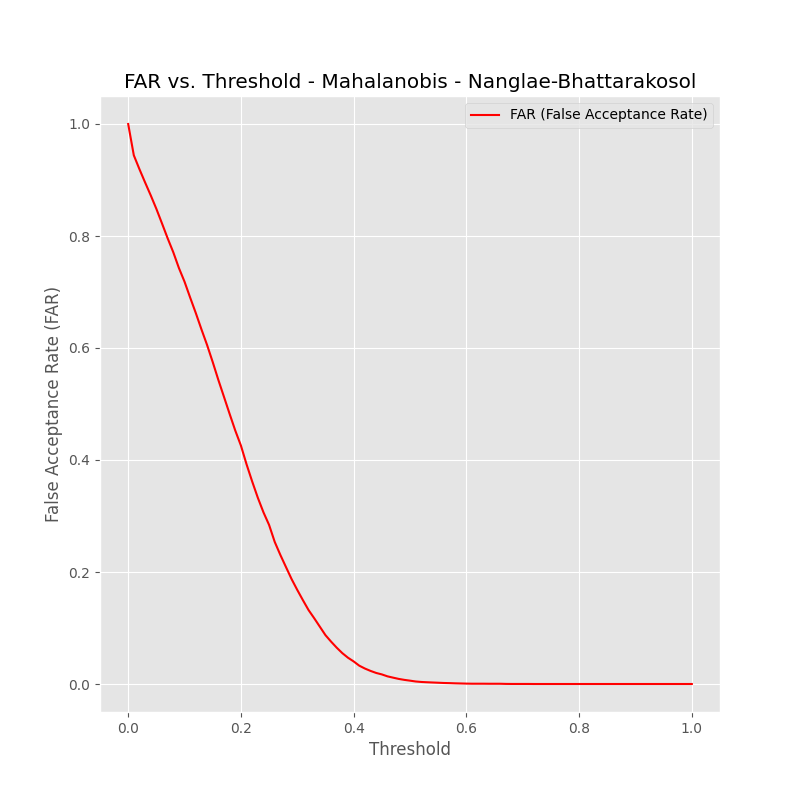}
    \end{minipage}
    \hfill
    \begin{minipage}{0.49\textwidth} 
        \centering
        \includegraphics[width=1.2\linewidth]{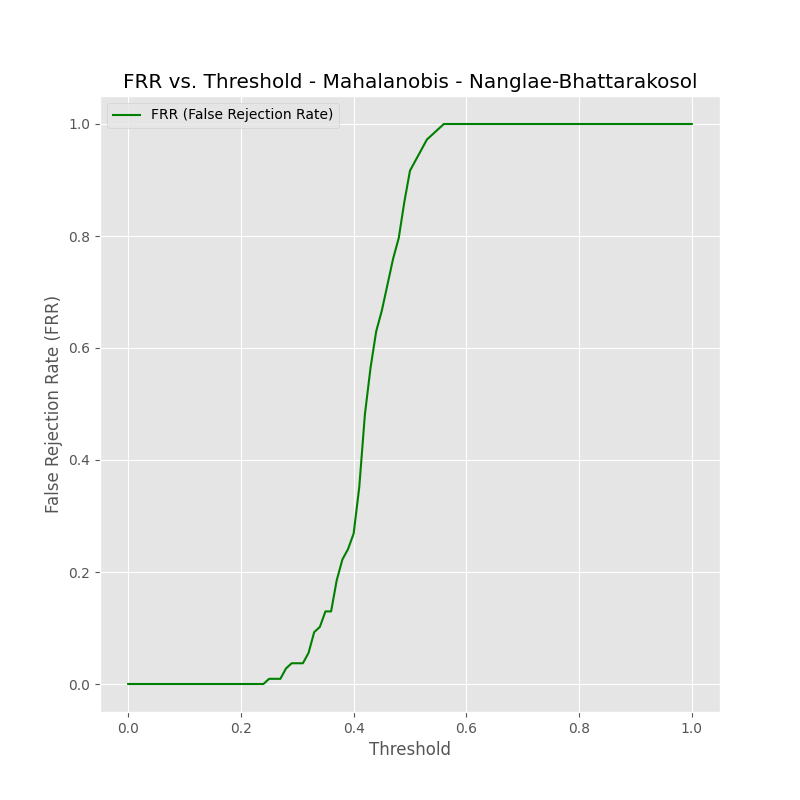}
    \end{minipage}
    \caption{\textbf{Nanglae-Bhattarakosol } AUC = 0.958, EER = 0.102}
\end{figure}
\FloatBarrier

\subsection{Gaussian Mixture Model}
Our implementation of GMM was directly inspired by an internet repository\cite{ref:gmm-code}.\\

The plots in Figure 6,7,8 and 9 make it clear that this model is the best out of the three we tried. It has excellent authentication performance, with an AUC of around 0.9 and an EER of around 0.15 on all datasets. Moreover the FAR and FRR plots show that GMM has a much more controlled behavior, as  when the threshold increases, the FAR decreases and the FRR increases. The drop-off as the threshold increases is more gradual. You can also see that performance is somewhat different on Aalto, but that is expected since it contains much more data than the other datasets.
\newpage

\begin{figure}[H]
    \centering
    \includegraphics[width=0.7\linewidth]{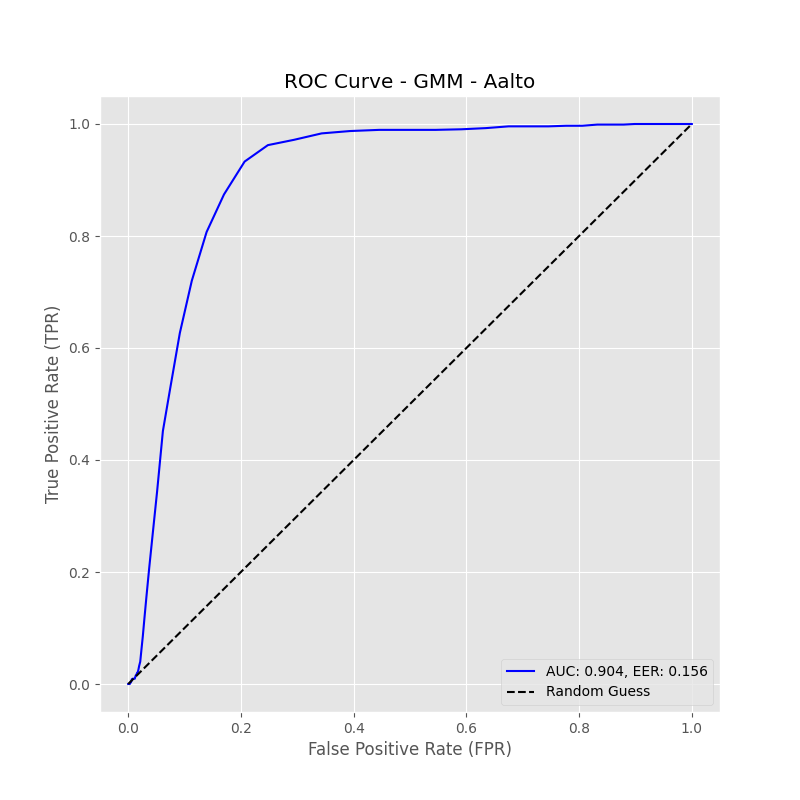}
\end{figure}
\begin{figure}[H]
    \centering
    \begin{minipage}{0.49\textwidth} 
        \centering
        \includegraphics[width=1.2\linewidth]{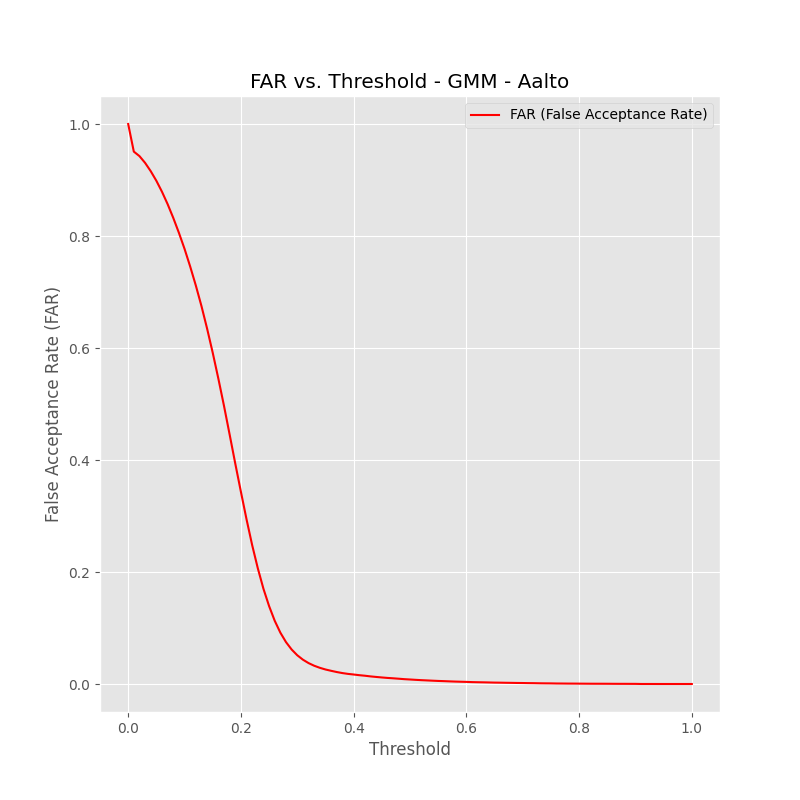}
    \end{minipage}
    \hfill
    \begin{minipage}{0.49\textwidth} 
        \centering
        \includegraphics[width=1.2\linewidth]{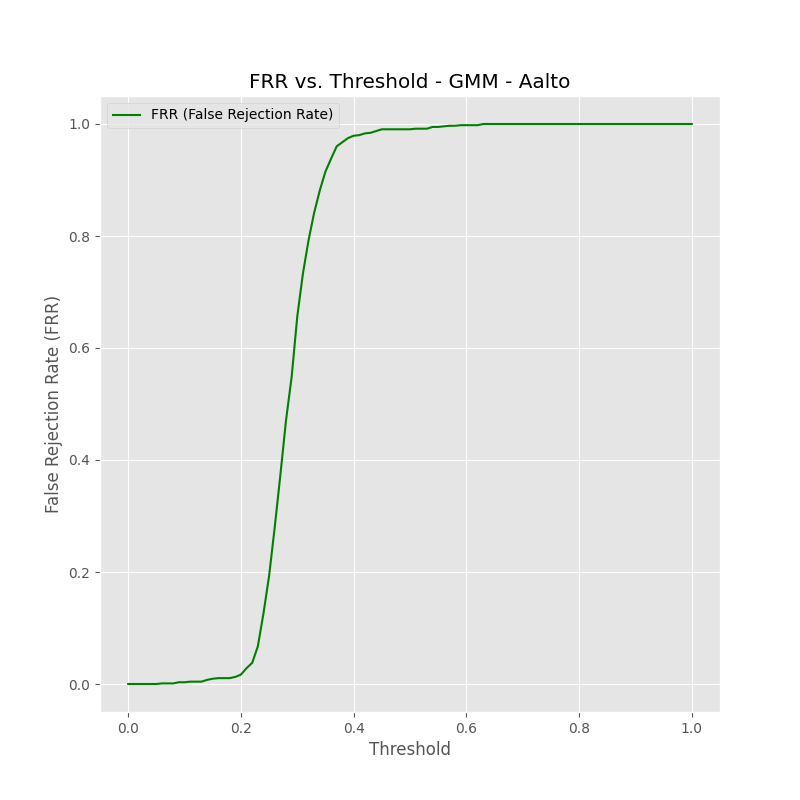}
    \end{minipage}
    \caption{\textbf{Aalto } AUC = 0.904, EER = 0.156}
\end{figure}

\begin{figure}[H]
    \centering
    \includegraphics[width=0.7\linewidth]{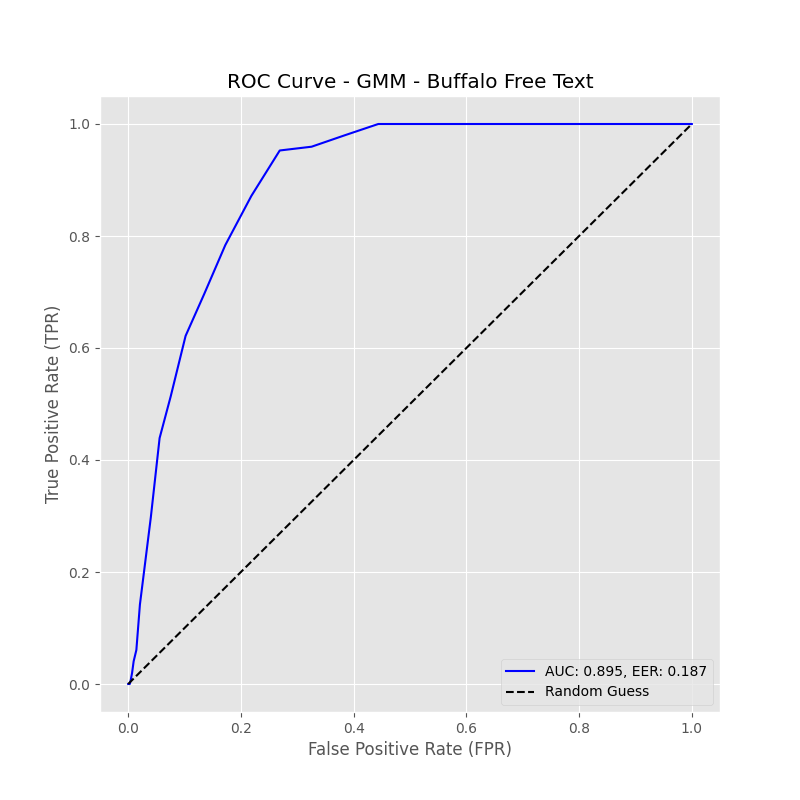}
\end{figure}
\begin{figure}[H]
    \centering
    \begin{minipage}{0.49\textwidth} 
        \centering
        \includegraphics[width=1.2\linewidth]{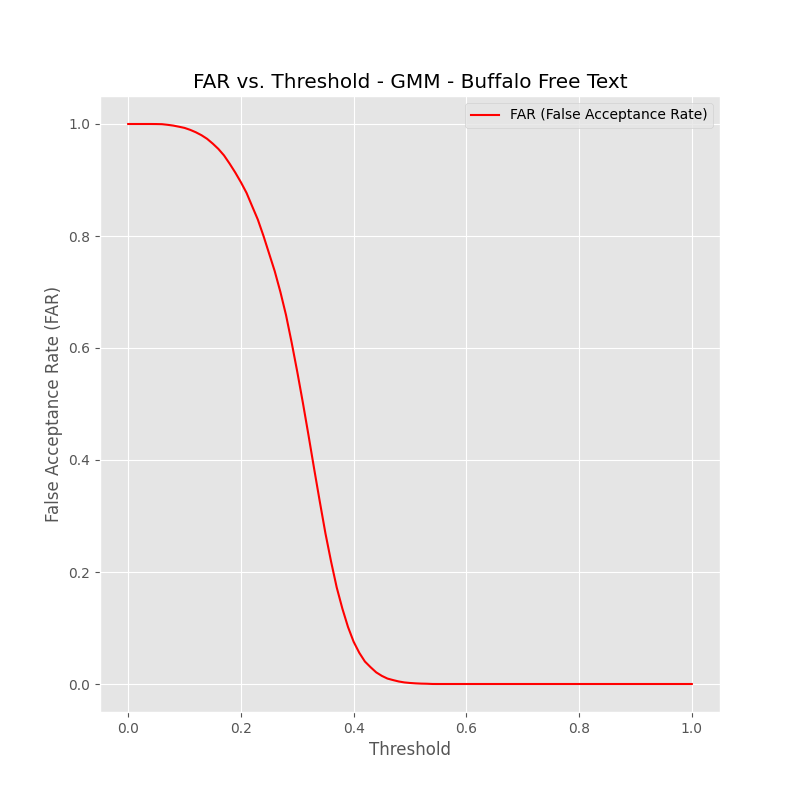}
    \end{minipage}
    \hfill
    \begin{minipage}{0.49\textwidth} 
        \centering
        \includegraphics[width=1.2\linewidth]{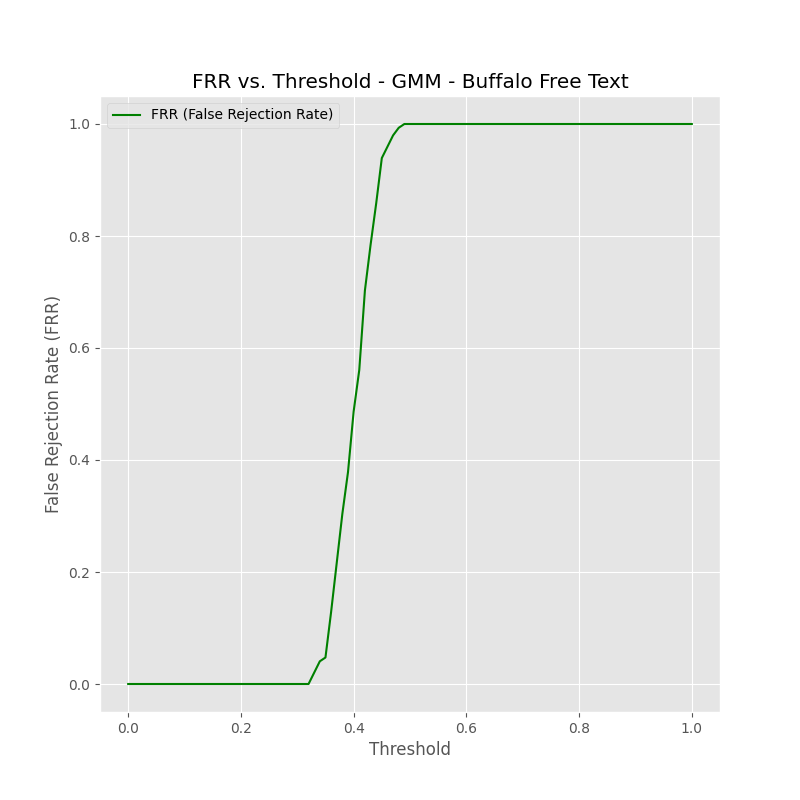}
    \end{minipage}
    \caption{\textbf{Buffalo Free-Text } AUC = 0.895, EER = 0.187}
\end{figure}

\begin{figure}[H]
    \centering
    \includegraphics[width=0.7\linewidth]{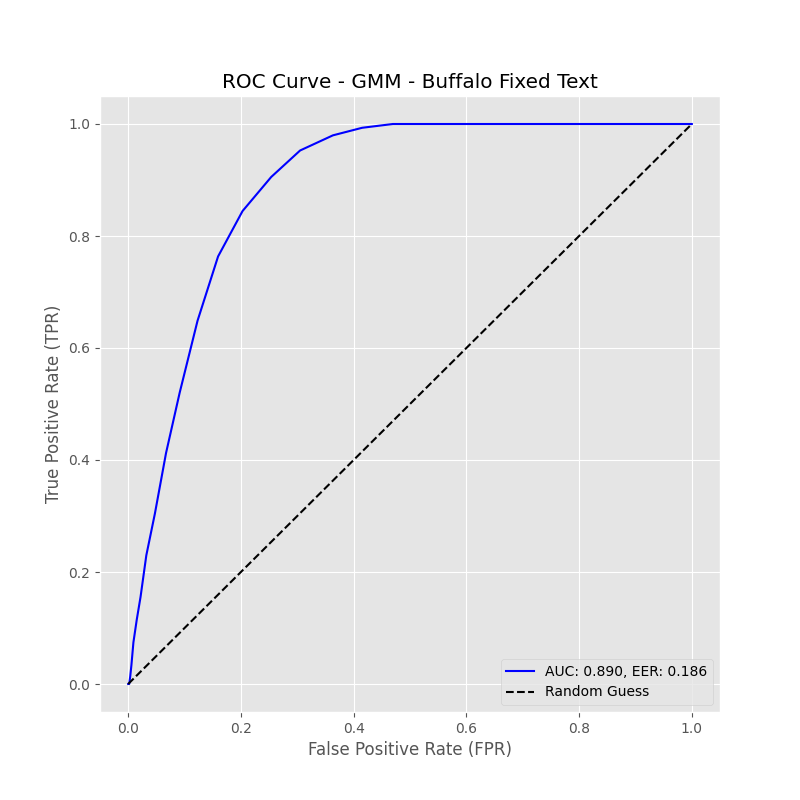}
\end{figure}
\begin{figure}[H]
    \centering
    \begin{minipage}{0.49\textwidth} 
        \centering
        \includegraphics[width=1.2\linewidth]{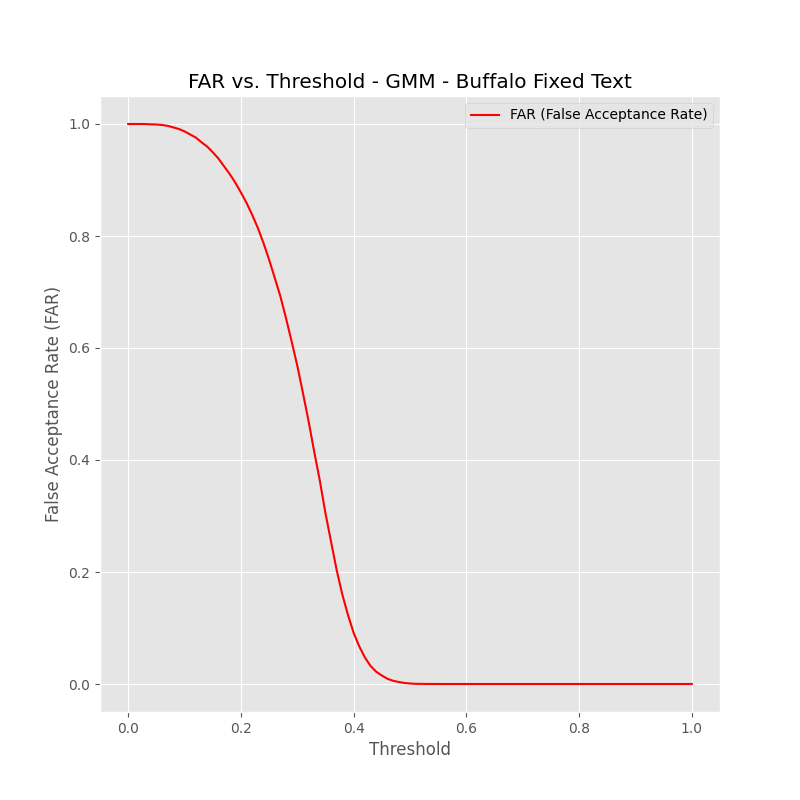}
    \end{minipage}
    \hfill
    \begin{minipage}{0.49\textwidth} 
        \centering
        \includegraphics[width=1.2\linewidth]{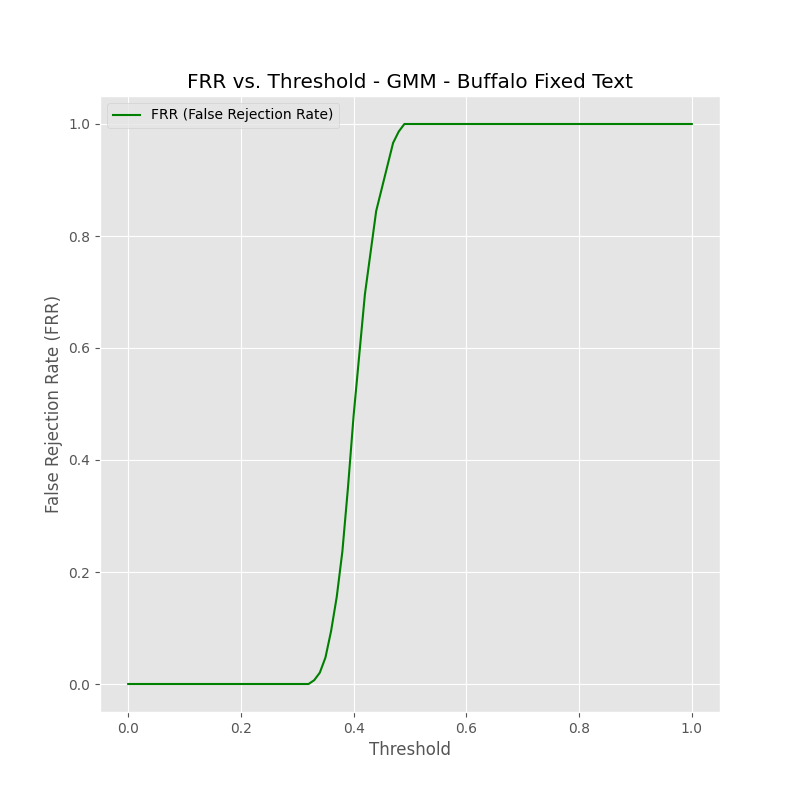}
    \end{minipage}
    \caption{\textbf{Buffalo Fixed-Text } AUC = 0.890, EER = 0.186}
\end{figure}

\begin{figure}[H]
    \centering
    \includegraphics[width=0.7\linewidth]{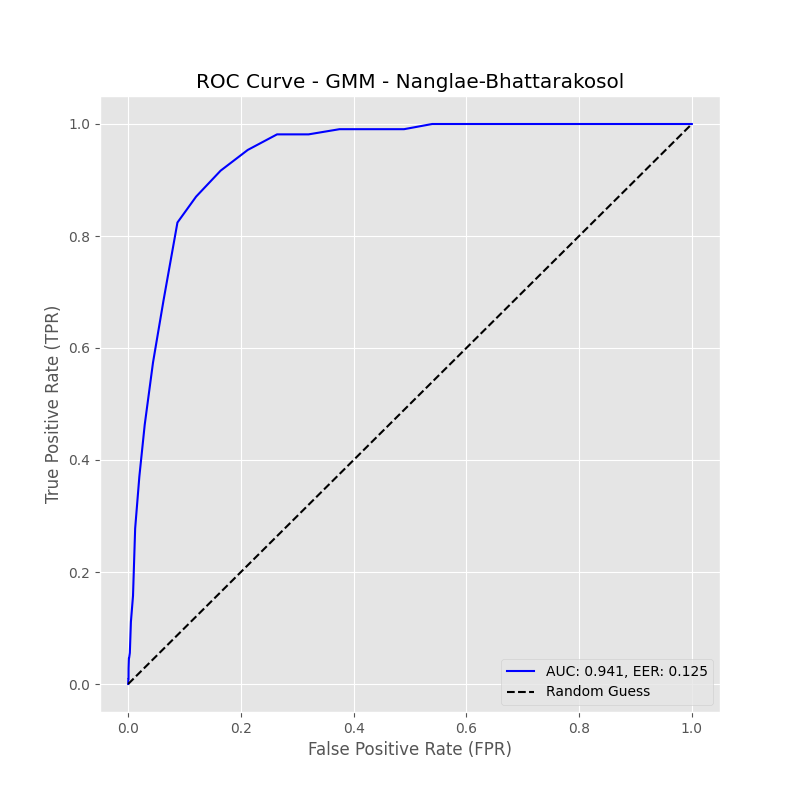}
\end{figure}
\begin{figure}[H]
    \centering
    \begin{minipage}{0.49\textwidth} 
        \centering
        \includegraphics[width=1.2\linewidth]{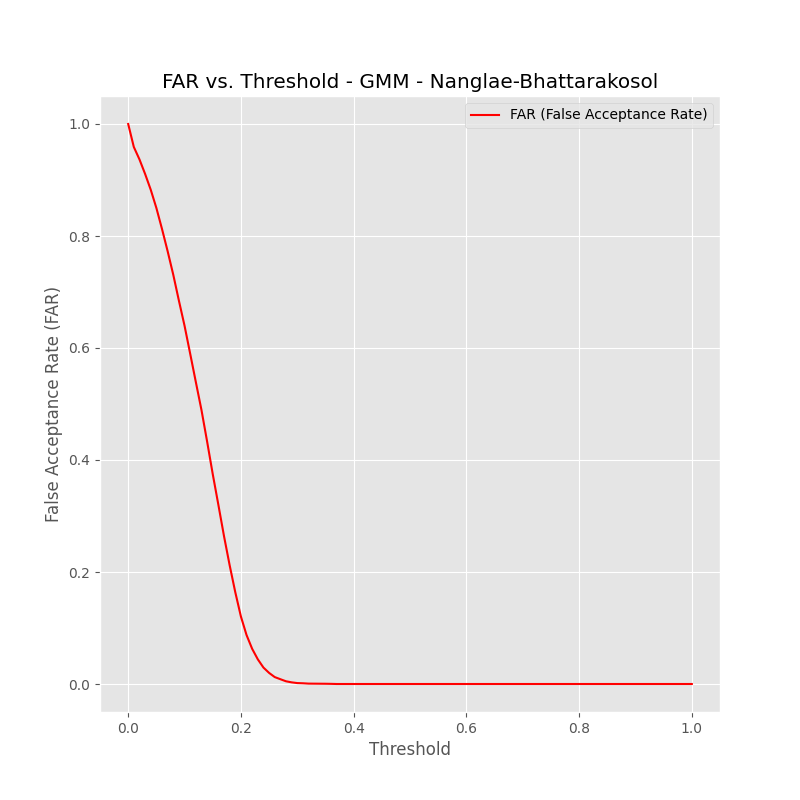}
    \end{minipage}
    \hfill
    \begin{minipage}{0.49\textwidth} 
        \centering
        \includegraphics[width=1.2\linewidth]{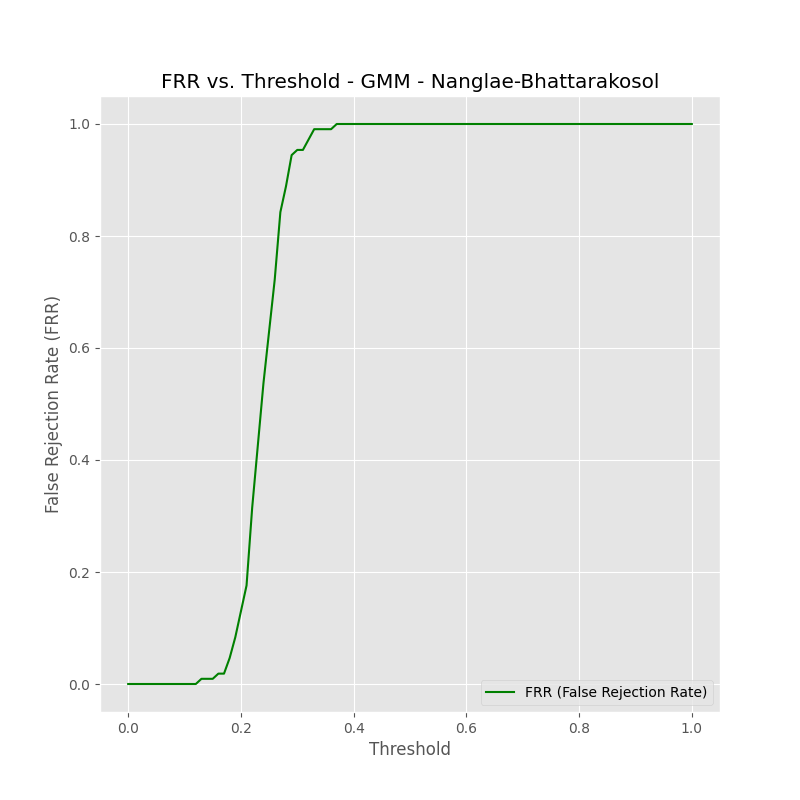}
    \end{minipage}
    \caption{\textbf{Nanglae-Bhattarakosol } AUC = 0.941, EER = 0.125}
\end{figure}
\FloatBarrier
\subsection{Gunetti Picardi}
We implemented the Gunetti-Picardi algorithm by looking at this implementation \cite{ref:gp-code}. We used the open set all-probe-against-all-gallery approach.\\

As we explained before, Gunetti Picardi is used primarily with free-text datasets, so we only used it on two sessions of the free-text section of Buffalo dataset. There are many possible combinations of A and R measures; however we choose to focus on these ones:
\begin{table}[H]
    \centering
    \begin{tabular}{|l|c|c|c|c|c|c|c|c|c|c} 
        \hline
         & $A_2$ & $A_3$ & $A_{23}$ & $R_2$ & $R_{23}$ & $R_2A_2$ & $R_2A_{234}$ & $R_{23}A_{23}$ & $R_{234}A_{23}$ \\
        \hline
        FAR & 13\% & 13\% & 13\% & 12\% & 10\% & 10\% & 10\% & 12\% & 8\% \\ 
        FRR & 13\% & 14\% & 13\% & 12\% & 11\% & 9\% & 9\% & 11\% & 8\% \\ 
        \hline
    \end{tabular}
    \caption{\textbf{Buffalo Free-Text}}
\end{table}
\FloatBarrier

\subsection{Conclusions}
As you can see, the AUC for all methods is similar, but Gaussian Mixture Model is by far the best method when it comes to EER. Across all datasets it always has the lowest EER. Curiously, you can also see that even with two different methods the FAR and FRR plots for the same datasets follow the same rough trajectory, this happens because while the methods may be different, data is still the most important element. Performance is always better when testing on the Nanglae-Bhattarakosol dataset. This is certainly caused by its small size, since when the same methods are used on the other datasets the performance is always lower. There is also a clear gap in performance when the same method is used on Aalto and when it is used on Buffalo. This might be explained by the fact that each Buffalo user has a lot more keystrokes samples compared to an Aalto user. This means that there are more variations per user.\\

You can verify our results independently by accessing our provided code \cite{ref:our-repo}.

\newpage